\documentclass[aps,prl,twocolumn,superscriptaddress,showpacs,amsmath,amssymb]{revtex4-1}
\usepackage{amsmath}
\usepackage{amsfonts}
\usepackage{amssymb}
\usepackage{graphicx}
\usepackage{color}

\usepackage{bm}
\usepackage{braket}
\usepackage{hyperref}

\begin{document}

\title{Topological $Z_2$ invariant in Kitaev spin liquids: \\
Classification of gapped spin liquids beyond projective symmetry group}

\author{Masahiko G. Yamada}
\email[]{myamada@mp.es.osaka-u.ac.jp}
\affiliation{Department of Materials Engineering Science, Osaka University, Toyonaka 560-8531, Japan}
\affiliation{Institute for Solid State Physics, University of Tokyo, Kashiwa 277-8581, Japan}

\date{\today}

\begin{abstract}
A projective symmetry group (PSG) has been regarded as a classification theory of
spin liquids.  However, it does not include a symmetry-protected topological order
of fermionic spinon excitations, and thus the classification of gapped spin liquids
is incomplete.  We demonstrate the classification beyond PSG by utilizing
the Kitaev model on the squareoctagon lattice, where two gapped spin liquids
are distinguished by a topological $Z_2$ invariant.
This $Z_2$ invariant can be defined solely by the time-reversal
and translation symmetries on condition that the time-reversal symmetry
is implemented projectively.  Thus, it is a hidden class of topological
Kitaev spin liquids with helical edge states, which has been ignored for a long time.
This suggests that there exists an unknown classification scheme of gapped spin liquids
beyond PSG.
\end{abstract}

\maketitle

\textit{Introduction}. ---
A projective symmetry group (PSG)~\cite{Wen2002PRB} has been regarded as a
classification theory of quantum spin liquids~\cite{Wen2002PLA}.  This theory is
based on the standard mean-field treatment of spin liquids.
In quantum spin liquids, the symmetries of the mean-field Hamiltonian
and the original spin model can be different.  The mean-field Hamiltonian
breaks the original symmetry and the symmetry is restored by
an additional gauge transformation.  The structure of the gauge transformation
is called PSG, which characterizes the spin liquid.

Despite its popularity, PSG does not include a 
topological insulator/superconductor property~\cite{Kane2005sep,Kane2005nov,Fu20073d,Fu2007inv}
of a fermionic spinon excitation (FSE).  FSE potentially possesses a symmetry-protected
topological (SPT) order~\cite{Yamada2017xsl}.
The existence of an SPT phase in FSE suggests the emergence of a novel class of
symmetry-enriched topological (SET) phases in the original spin model.  The PSG theory
is incomplete as it cannot classify this class of SET phases where FSE has an SPT order~\cite{Qi2015,Lu2017}.
In fact, PSG depends on the mean-field approximation of spin liquids~\cite{Essin2013},
which in principle is impossible to treat the property of FSEs directly~\cite{Kitaev2006}.
We extend the topological transition between two gapped spin liquids with the same
PSG, and exactly show that it is a topological transition between two different SET phases
by employing the Kitaev model on the squareoctagon lattice.

However, PSG still plays an important role in the classification of these phases.
This is because it is also impossible for this classification beyond PSG to be described
by topological periodic tables.  Conventionally, Kitaev models including the one
on the honeycomb lattice~\cite{Kitaev2006} with a time reversal symmetry are regarded as
class BDI in the topological periodic table, which is trivial in two dimensions and
cannot host an SPT order~\cite{Kitaev2009,Ryu2010}.
Thus, naively there is no topological $Z_2$ invariant in Kitaev spin liquids by this
free-fermionic classification.  On the other hand, the situation is different in the case of
the squareoctagon lattice~\cite{Yang2007,Baskaran2009,Kells2011}.
This is because the action of the time-reversal symmetry
is nonlocal in the $k$-space~\cite{Hermanns2014,Hermanns2015}, which potentially makes
the conventional topological periodic table classification inapplicable.
This exotic PSG may allow another classification of topological $Z_2$ invariants
inside the same class of PSG thanks to the projective nature of symmetries.

It is important to note that the nontriviality of PSG in the squareoctagon model
does not automatically mean that Majorana fermions host a nontrivial SPT order.
There is still an additional topological $Z_2$ invariant not included in the PSG classification.
Indeed, the Kitaev model on the squareoctagon lattice has two gapped phases separated
by the gapless line.  While these two phases have the same PSG and are indistinguishable
in the conventional scheme, we define a $Z_2$ invariant which can distinguish between
them and demonstrate the topological phase transition between these two phases.

Our use of the Kitaev model is advantageous to the Heisenberg model.
In the Kitaev model, the ground state is exactly solvable and the $\pi$-flux
state is guaranteed to be stabilized by Lieb's theorem~\cite{Lieb1994}.
In the case of the Heisenberg model, the $\pi$-flux state is just one of many
mean-field solutions, and the brute-force approach is necessary in the discussion.
In addition, the geometry of the squareoctagon lattice is also an important
ingredient for the existence of a nontrivial $Z_2$ invariant.

The definition of the $Z_2$ invariant itself is exotic and we call it
``nonlocal Pfaffian invariant''.  The discussion follows Fu and Kane~\cite{Fu2007inv},
but the definition is slightly different from theirs, reflecting the nontrivial PSG.
Due to the nonlocal property of the time reversal, the
Fourier-transformed form has an additional $k$-space translation in the first Brillouin zone.
The original Fu-Kane invariant does not work due to this momentum shift, and a new
invariant has to be defined nonlocally in the $k$-space.

Indeed, the topological $Z_2$ invariant can be defined solely by the time-reversal
and translation symmetries in 2 or 3 dimensions, and thus it is a hidden class of
topological Kitaev spin liquids, which has been ignored in the previous classification
of Kitaev spin liquids~\cite{Obrien2016}.
The classification is not only beyond PSG but also beyond the topological periodic
table, which assumes no momentum shift for the time-reversal symmetry.
Indeed, a $Z_2$ nontrivial phase can only exist in the case that the time-reversal
symmetry accompanies a momentum shift, and thus the strong correlation in the Kitaev
model is a necessary condition.
The additional $k$-space translation makes it possible to have a $Z_2$ invariant
beyond the previous class BDI classification.

In this Rapid Communication, we define the topological $Z_2$ invariant protected by the time-reversal
symmetry, and discover edge states in the $Z_2$ nontrivial
phase.  Our theory suggests the existence of a huge number of ignored SET phases described
by FSEs with SPT order by extending our results to topological crystalline phases~\cite{Fu2011}.

\textit{Lattice}. ---
The Kitaev model, originally defined for the honeycomb lattice~\cite{Kitaev2006},
has bond-dependent anisotropic interactions.  Usually bonds on the lattice is colored by
three different colors, red, green, and blue.  Red, green, and blue bonds have $x$-, $y$-,
and $z$-directional Ising interactions, respectively, and the ground state becomes a spin
liquid for any tricoordinated lattices, as long as each site is connected to three different
types of bonds and the ground state flux sector is symmetric.  This condition holds for both
the honeycomb and squareoctagon lattices. The Hamiltonian is
\begin{equation}
    H=-J_x\sum_{\langle jk\rangle \in x} \sigma_j^x \sigma_k^x-J_y\sum_{\langle jk\rangle \in y} \sigma_j^y \sigma_k^y-J_z\sum_{\langle jk\rangle \in z} \sigma_j^z \sigma_k^z,
\end{equation}
where $\bm{\sigma}_j$ are Pauli matrices, and $\langle jk\rangle \in x$, $y$, and $z$ are
red, green, and blue bonds, respectively.

\begin{figure}
\centering
\includegraphics[width=8.6cm]{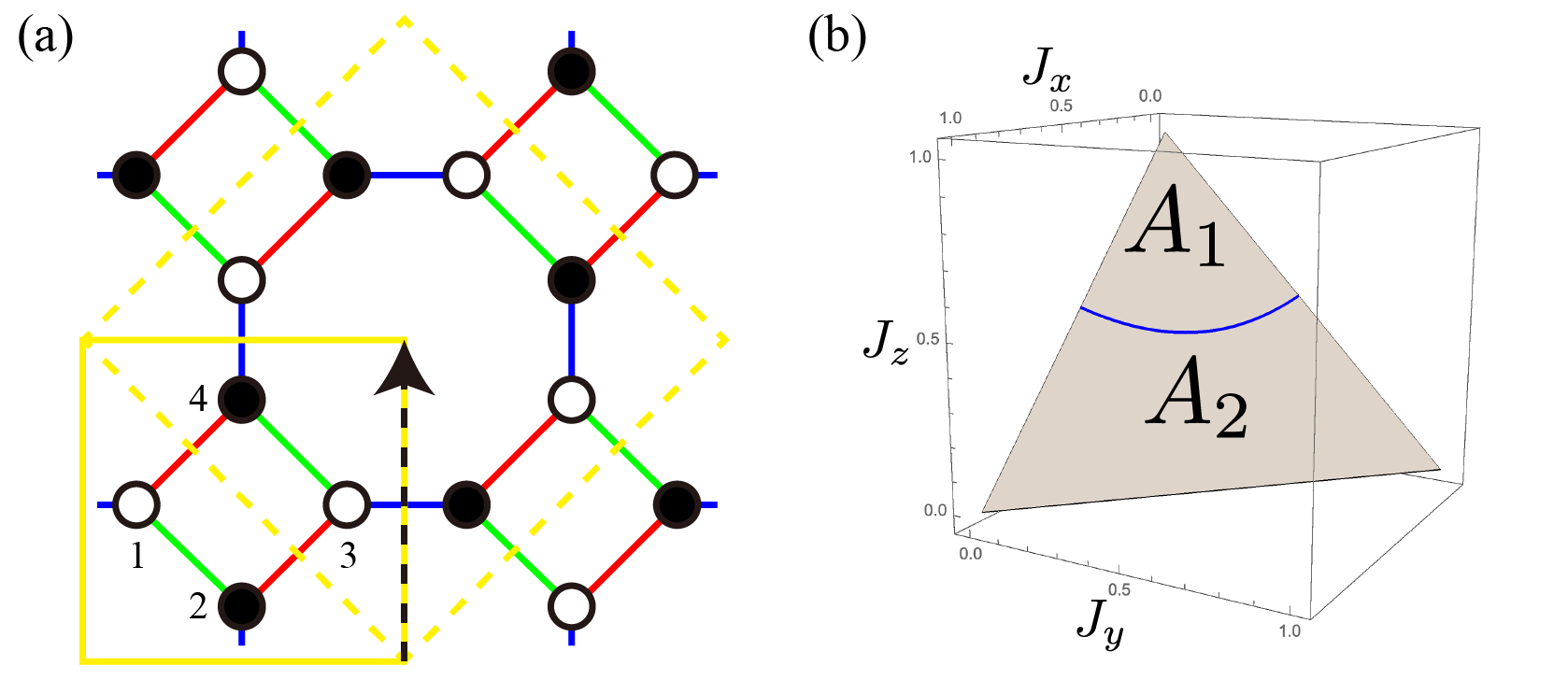}
\caption{(a) Kitaev model on the squareoctagon lattice.  Red, green, and blue bonds
have $x$-, $y$-, and $z$-directional Ising interaction, respectively, and odd and even
sublattices are distinguished by white and black circles, respectively.
(b) Phase diagram on the plane $J_x+J_y+J_z = 1$~\cite{Yang2007}.
$A_1$ phase is a topological one with edge states, while $A_2$ phase is not.}
\label{lattice}
\end{figure}

This model can be solved by introducing Majorana fermions~\cite{Kitaev2006}.
We would not follow the detailed description of how to solve this model,
but eventually the problem of solving this spin model is reduced to a free-fermion model
coupled to a ``magnetic'' field called flux with a single Majorana mode $c_j$.
The Hamiltonian can always be recast to this form:
\begin{equation}
    H_\textrm{Majorana} = \frac{i}{4} \sum_{j,k} A_{jk} c_j c_k,\label{Majorana}
\end{equation}
where $A$ is a real skew-symmetric matrix with the following property.
\begin{equation}
    A_{jk}=
    \begin{cases}
        \pm 2J_\gamma & \langle jk \rangle \in \gamma \\
        0 & \textrm{otherwise}
    \end{cases},
\end{equation}
where the sign is determined as follows.  For solid bonds in Fig.~S1 in Supplemental Material,
$+$ when $j$ is on the even sublattice and $k$ is on the odd sublattice, while $-$ when
opposite.  For dashed bonds in Fig.~S1 in Supplemental Material (SM),
$-$ when $j$ is on the even sublattice and $k$ is on the odd sublattice, while $+$ when
opposite~\footnote{See Supplemental Material at [URL will be inserted by publisher] for Fig.~S1
and the PSG classification.}.
By diagonalizing this Hamiltonian, we can get the ground-state spectrum and all SPT
information is included.

Kitaev models on the two-dimensional (2D) squareoctagon lattice~\cite{Yang2007,Baskaran2009,Kells2011}
can actually be defined in many ways, but we use the most symmetric coloring by
respecting the translation symmetry of the skeletal structure [see Fig.~\ref{lattice}(a)].
In this case, every bond in the same direction has the same color.
Lieb's theorem is applicable and the ground state sector is $\pi$-flux~\cite{Lieb1994}.
Thus, we only consider a $\pi$-flux state in this Rapid Communication.

It was found that there are two gapped phases $A_1$ and $A_2$
in the Kitaev model on the squareoctagon lattice [see Fig.~\ref{lattice}(b)]~\cite{Yang2007}.
It is important to note that the flux sector and PSG do not change between
these two phases, so these two gapped phases are indistinguishable by PSG.
These phases are separated by the gapless line with two Dirac cones at
$(0,0)$ and $(\pi,\pi)$ in the usual Brillouin zone.  The phase boundary can be
written as $J_x^2 + J_y^2 = J_z^2$, so $J_x=J_y=1/4,$ $J_z=1/2$ is in $A_1$ phase,
while $J_x=J_y=J_z=1/3$ is in $A_2$ phase~\cite{Yang2007}.  The existence of
Dirac cones suggests that two phases are distinguished by some topological number.

\begin{figure}
\centering
\includegraphics[width=4.3cm]{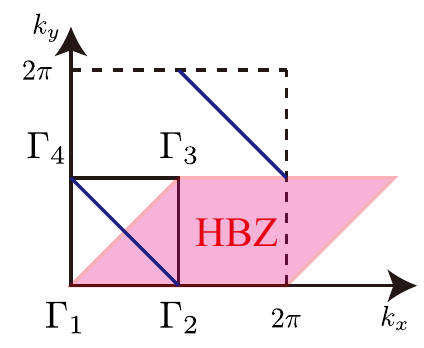}
\caption{Inversion invariant momenta $\Gamma_i$ ($i=1,\dots4$) are shown.  A half Brillouin
zone (HBZ) is represented by the pink shaded region.  Blue solid lines represent the subsystem
where the DIII topological number is defined.}
\label{hbz}
\end{figure}

\textit{Projective symmetry}. ---
Since a $Z_2$ gauge is fixed in Eq.~\eqref{Majorana}, the symmetry operation can
potentially change the gauge in Majorana fermion systems.  Differently from electronic
systems, where symmetry actions do not accompany the gauge change, symmetries in Majorana
systems often require an additional gauge transformation in order to act within the
fixed gauge sector.  Such a supplemental gauge transformation sometimes leads
to a momentum shift in the $k$-space.  Following Refs.~\cite{Kitaev2006,Hermanns2014,Obrien2016},
we review the properties of (projective) time-reversal and inversion symmetries
in the Kitaev model.

Indeed, the implementation of the most fundamental symmetry operation, time reversal,
is exotic with a momentum shift, which changes the topological classification.
According to PSG, the time-reversal operator accompanies a gauge transformation
which is defined by a sublattice parity $(-1)^j$ for the $j$th site.
This fact can be understood as follows.  Naively, a candidate time-reversal symmetry
is defined by a complex conjugation $K$ as $K c_j K = c_j$.  However,
$K$ changes the sign of a hopping term $ic_j c_k$, and thus needs to be supplemented
by some gauge transformation.  In bipartite lattices, the simplest choice of the
gauge transformation is the sublattice parity $(-1)^j$, which is 1 for the even sublattice
and $-1$ for the odd sublattice.  Since for each hopping term $j$ and $k$ lie on different
sublattices, this gauge transformation always supplements the sign caused by the complex
conjugation.

Therefore, the time reversal operation can be written as $\Theta = (-1)^j K$.
This fact causes an interesting effect on the squareoctagon lattice.
Since in the squareoctagon lattice the sublattice parity is not commensurate
with the translation symmetry [see Fig.~\ref{lattice}(a)], the unit cell
shown by the solid yellow line is effectively enlarged to the dashed yellow line.
If we use the original unit cell, an additional $k$-space translation with
a wavevector $\bm{k}_0 = (\pi,\pi)$ is necessary after the Fourier transformation.
Thus, due to the projectiveness the action of the time reversal in the $k$-space
becomes between $\bm{k}$ and $\bm{k}_0 - \bm{k}$~\cite{Hermanns2014}.  This
exotic nature of time reversal is relevant to the classification of SPT phases.
The topological classification becomes different from the usual class BDI, resulting in the
possibility of the existence of an exotic phase impossible in the classification
based on the topological periodic table.

As for the (bond-centered) inversion symmetry, the gauge transformation does not
enlarge the unit cell.  The inversion is supplemented by the following
gauge transformation $G_I$ in the $k$-space.  The site index
from 1 to 4 for the following matrices is shown in Fig.~\ref{lattice}(a).
\begin{equation}
    G_I = \begin{pmatrix}
    -1 & 0 & 0 & 0 \\
    0 & 1 & 0 & 0 \\
    0 & 0 & 1 & 0 \\
    0 & 0 & 0 & -1
\end{pmatrix},
\end{equation}
with a site change
\begin{equation}
    P_\textrm{phys} = \begin{pmatrix}
    0 & 0 & 1 & 0 \\
    0 & 0 & 0 & 1 \\
    1 & 0 & 0 & 0 \\
    0 & 1 & 0 & 0
\end{pmatrix},
\end{equation}
Thus, the total operator $P$ for the inversion in the $k$-space is
\begin{equation}
    P = G_I P_\textrm{phys} = \begin{pmatrix}
    0 & 0 & -1 & 0 \\
    0 & 0 & 0 & 1 \\
    1 & 0 & 0 & 0 \\
    0 & -1 & 0 & 0
\end{pmatrix},
\end{equation}
While the inversion $P$ connects $\bm{k}$ and $-\bm{k}$ in the $k$-space as usual,
it holds that $P^2 = -1$.

\textit{Symmetry indicator approach}. ---
We quickly construct a Fu-Kane symmetry indicator~\cite{Fu2007inv}
which distinguishes the two phases.  The existence of the symmetry indicator automatically
proves that the two phases are separated as long as the inversion symmetry is protected,
but later we will find that the inversion symmetry is not necessary and actually
the same invariant can be defined solely by the exotic time-reversal symmetry.

Assuming the inversion symmetry $P$ with $P^2 = -1$,
the inversion eigenvalues $\pm i$ can be defined at every inversion-invariant momentum (IIM),
$\Gamma_1 = (0,0)$, $\Gamma_2 = (\pi,0)$, $\Gamma_3 = (\pi,\pi)$, and $\Gamma_4 = (0,\pi)$
[see Fig.~\ref{hbz}].
We note that these IIM are not time-reversal-invariant, and thus here they
are not called time-reversal-invariant momenta.

Since $[\Theta,P]=0$ and $\Theta$ is antiunitary, the time reversal
flips the sign of the inversion eigenvalue.
If we define the inversion eigenvalue of the $\alpha$th Bloch
eigenstate $\ket{u_{\alpha,\bm{k}=\Gamma_i}}$ as $i\xi_{\alpha}(\Gamma_i)$
with $\xi_{\alpha}(\Gamma_i) = \pm 1$ ($i=1,\dots,4$), $\Gamma_1$ and $\Gamma_3$
($\Gamma_2$ and $\Gamma_4$) are always related with each other.  In fact,
$\xi_{\alpha}(\Gamma_1) = -\xi_{\alpha}(\Gamma_3)$ and $\xi_{\alpha}(\Gamma_2) = -\xi_{\alpha}(\Gamma_4)$,
so the $Z_2$ invariant should be defined only for $\Gamma_1$ and $\Gamma_2$ to erase
the redundancy.  A candidate $Z_2$ invariant to distinguish two phases is
\begin{equation}
    \delta = \prod_{\alpha=1}^{N} \prod_{i=1}^2 \xi_{\alpha}(\Gamma_i),
\end{equation}
where $N = 2$ is the number of occupied bands.

By checking the inversion eigenvalues, $A_1$ phase indeed has $\delta = -1$ and
topologically non-trivial, while $A_2$ phase has $\delta = 1$.  Thus, from this
simple guess, we conclude that $A_1$ phase is topological, \textit{i.e.} a new SET
phase, and $A_2$ phase is not.  This $Z_2$ invariant $\delta$ is related to a nonlocal
Pfaffian invariant, as discussed in SM.

\textit{Dimensional reduction approach}. ---
In the previous section, we intensively used the inversion symmetry to define
the $Z_2$ invariant.  However, the inversion symmetry is actually unnecessary,
and we seek a definition which will not require the inversion eigenvalues in this
section.

Another way of evaluating this topological number is to think that the time-reversal symmetry
effectively breaks the translation symmetry, and the unit cell is enlarged by this
``weak symmetry breaking''~\cite{Kitaev2006}.  In this enlarged unit cell, a more explicit
construction of the $Z_2$ invariant is possible.

This picture is indeed similar to the one used in type IV magnetic space
group~\cite{Watanabe2018}, which is useful to connect the $Z_2$ invariant
above to the previously discovered phases of topological insulators/superconductors.
It is closely related to antiferromagnetic topological insulators~\cite{Mong2010,Fang2013}
and we can follow their theory to construct a similar invariant in the enlarged
unit cell.

In the enlarged unit cell, the time-reversal symmetry is no longer projective,
but rather we need an additional ``half translation'' to restore the original
crystalline symmetry.  Since the time reversal in the Kitaev model accompanies the
gauge transformation which flips signs for one of the sublattices,
this sign flip is restored only by the half translation which switches the sublattice parity.
There are many half translation vectors, but we choose $\bm{D}=(1/2,1/2)$, as shown in
the black dashed arrow in Fig.~\ref{lattice}(a), for simplicity.
These two operations have the same gauge transformation, so the combination
of the two, denoted by $\Theta_S$, can be written without a gauge transformation
as $\Theta_S = T_{\bm{D}} K$,
where $T_{\bm{D}}$ is a half translation without a gauge transformation.

In the $k$-space, $\Theta_S$ is a new symmetry defined for the whole Brillouin zone.
Especially, $\Theta_S^2 = e^{2\bm{D}\cdot \bm{k}i}$, and $\Theta_S^2 = -1$
when $2\bm{D}\cdot \bm{k} = \pi$.  Thus, we have a Kramers degeneracy on the line
$2\bm{D}\cdot \bm{k} = \pi$, passing $(0,\pi)$ and $(\pi,0)$.
On this one-dimensional (1D) subsystem, shown as blue solid lines in Fig.~\ref{hbz},
$\Theta_S$ serves as an effective time-reversal symmetry of the 1D class DIII~\cite{Fang2013}.
Following the definition of the $Z_2$ invariant for the 1D topological
superconductor in class DIII~\cite{Budich2013diii},
the $Z_2$ invariant is again defined by a formula
Eq.~\eqref{diii} within the approach using an enlarged unit cell.

First, we define $\{\ket{\alpha}\}_\alpha$ as occupied Bloch functions at $(0,\pi)$, and
$\{\ket{\tilde{\alpha}}\}_\alpha$ as occupied Bloch functions at $(\pi,0)$.  We note that
the basis choice for occupied states is arbitrary at each $k$-point.  Then, we define matrices
$\big(w_S(0,\pi)\big)_{\alpha \beta} = \braket{\alpha|\Theta_S|\beta}$ and
$\big(w_S(\pi,0)\big)_{\alpha \beta} = \braket{\tilde{\alpha}|\Theta_S|\tilde{\beta}}$.
Intermediate states are also necessary to restore the basis independence as follows.
\begin{align}\label{diii}
    \delta &= \left( \det U^K \right)\frac{\textrm{Pf}[w_S(0,\pi)]}{\textrm{Pf}[w_S(\pi,0)]}. \\
    U^K_{\alpha\beta} &= \braket{\tilde{\alpha}|\left(\lim_{n\to\infty} \prod_{j=0}^n P_F(\bm{k}_j)\right)|\beta},
\end{align}
where $\bm{k}_j = (j\pi/n,\pi-j\pi/n)$ and $P_F(\bm{k})$ is a spectral projector onto
the occupied states at $\bm{k}$~\cite{Budich2013diii}.

If we compute the quantity above, indeed we reproduce $\delta=-1$ for $A_1$ phase
and $\delta=1$ for $A_2$ phase.  In the calculation we used $n=1000$ points between
$(0,\pi)$ and $(\pi,0)$.  This formula is useful not only because it is gauge-invariant
but also because it only requires the information of the time-reversal symmetry.
This proves that the inversion symmetry is not necessary for the phase protection.
As we have defined the same invariant in two different ways, we will move on to
its topological consequence, edge states~\cite{Wang2017}.

\begin{figure}
\centering
\includegraphics[width=8.6cm]{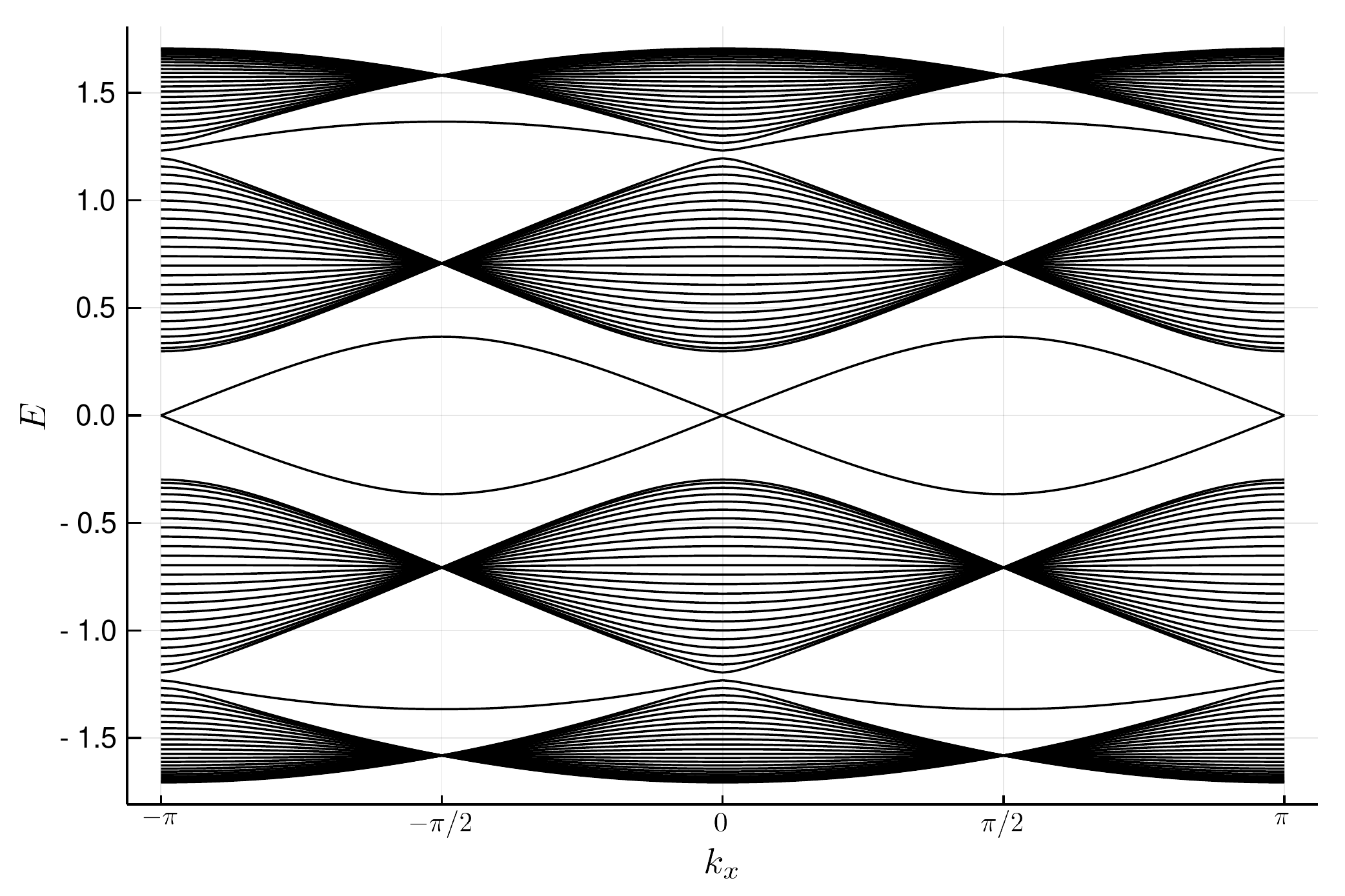}
\caption{Band structure for a 100-site strip of the squareoctagon lattice.
Energy $E$ is shown in the unit of $A_{jk}$.  $J_x=J_y=1/4$ and $J_z=1/2$,
deep inside $A_1$ phase, were used for the calculation.}
\label{edge}
\end{figure}

\textit{Edge states}. ---
It is a well-known fact in the case of antiferromagnetic topological
insulators that edge states exist for the nontrivial phase when the surface
is terminated in the $\Theta_S$-symmetric way~\cite{Fang2013,Shiozaki2016}.
This condition is met for almost every boundary termination for the squareoctagon lattice.
Therefore, we check the existence of edge states with the simplest open
boundary condition.

We used a strip geometry by simply repeating the original unit cell of the
squareoctagon lattice for 25 times along the $y$-direction.  The band structure
based on this termination is shown in Fig.~\ref{edge}.
The helical gapless points correspond to edge states and protected by the
time-reversal and particle-hole symmetries.  We note that each gapless point
is pinned at $k_x = 0$ or $k_x = \pi$ because of the time-reversal symmetry.

The meaning of the robust helical edge state cannot be captured within the
previous framework of SET phases because both $A_1$ and $A_2$ phases are
$Z_2$ topologically ordered states described by the toric code.
What makes difference is the topological band structure of FSEs, and its
protection by the time-reversal symmetry.  The general classification of
such phases is an interesting future problem.

\textit{Discussions}. ---
Here we defined a topological $Z_2$ invariant for Kitaev models in a certain class,
where the time-reversal symmetry accompanies a momentum shift.  We gave a definition
of the same invariant threefold (including the one in SM),
and discovered that the nontrivial phase is
protected solely by the time-reversal symmetry.  We only discussed the 2D system,
but the generalization to three dimensions is straightforward because the classification
of topological superconductors in class DIII of the 2D subsystem is again $Z_2$.
Necessary ingredients are the same in both 2D and 3D, and we need a momentum shift
in the $k$-space for the time-reversal symmetry.  A $\pi$ flux may also be useful
to have an unusual phase.
We note the disorder effect in such gapped phases is also an interesting
problem~\cite{Morimoto2015,Yamada2020}.

We can easily imagine another ``beyond PSG'' phase, a topological crystalline
spin liquid, protected by the crystalline symmetry~\cite{Yamada2017xsl,Yamada2018}.
The (8,3)-$n$ lattice is known to have a similar transition between two gapped phases,
but the time-reversal symmetry does not accompany a momentum shift~\cite{Obrien2016}.
Thus, the Kitaev model on this lattice is a potential candidate for a topological
crystalline spin liquid~\cite{Dwivedi2018,Zhao2020}.  Discovering topological (crystalline) insulators in tricoordinated
lattices realizable in iridates or Ru-compounds~\cite{Jackeli2009,Plumb2014,Yamada2017mof}
is also a future problem.

The conclusion implies that PSG does not give a complete classification
of gapped spin liquids.  We would need a post-PSG theory which can
correctly distinguish the SPT orders of FSEs in order to classify
gapped spin liquids correctly in the future.

\begin{acknowledgments}
We thank V.~Dwivedi, D.~Else, L.~Fu, S.~Fujimoto, M.~Hermanns, C.~Hickey, H.~Katsura, T.~Morimoto,
S.~Ono, H.~C.~Po, K.~Shiozaki, and Y.~Tada.
A part of this work has been done during the visit to Massachusetts Institute of Technology,
supported by JSPS.
M.G.Y. is supported by the Materials Education program for the future leaders in Research, Industry, and Technology (MERIT), and by JSPS.
This work was supported by JST CREST Grant Number JPMJCR19T5, Japan,
and by JSPS KAKENHI Grant Numbers JP17J05736.
\end{acknowledgments}

\bibliography{paper}

\begin{thebibliography}{36}%
\makeatletter
\providecommand \@ifxundefined [1]{%
 \@ifx{#1\undefined}
}%
\providecommand \@ifnum [1]{%
 \ifnum #1\expandafter \@firstoftwo
 \else \expandafter \@secondoftwo
 \fi
}%
\providecommand \@ifx [1]{%
 \ifx #1\expandafter \@firstoftwo
 \else \expandafter \@secondoftwo
 \fi
}%
\providecommand \natexlab [1]{#1}%
\providecommand \enquote  [1]{``#1''}%
\providecommand \bibnamefont  [1]{#1}%
\providecommand \bibfnamefont [1]{#1}%
\providecommand \citenamefont [1]{#1}%
\providecommand \href@noop [0]{\@secondoftwo}%
\providecommand \href [0]{\begingroup \@sanitize@url \@href}%
\providecommand \@href[1]{\@@startlink{#1}\@@href}%
\providecommand \@@href[1]{\endgroup#1\@@endlink}%
\providecommand \@sanitize@url [0]{\catcode `\\12\catcode `\$12\catcode
  `\&12\catcode `\#12\catcode `\^12\catcode `\_12\catcode `\%12\relax}%
\providecommand \@@startlink[1]{}%
\providecommand \@@endlink[0]{}%
\providecommand \url  [0]{\begingroup\@sanitize@url \@url }%
\providecommand \@url [1]{\endgroup\@href {#1}{\urlprefix }}%
\providecommand \urlprefix  [0]{URL }%
\providecommand \Eprint [0]{\href }%
\providecommand \doibase [0]{http://dx.doi.org/}%
\providecommand \selectlanguage [0]{\@gobble}%
\providecommand \bibinfo  [0]{\@secondoftwo}%
\providecommand \bibfield  [0]{\@secondoftwo}%
\providecommand \translation [1]{[#1]}%
\providecommand \BibitemOpen [0]{}%
\providecommand \bibitemStop [0]{}%
\providecommand \bibitemNoStop [0]{.\EOS\space}%
\providecommand \EOS [0]{\spacefactor3000\relax}%
\providecommand \BibitemShut  [1]{\csname bibitem#1\endcsname}%
\let\auto@bib@innerbib\@empty
\bibitem [{\citenamefont {Wen}(2002{\natexlab{a}})}]{Wen2002PRB}%
  \BibitemOpen
  \bibfield  {author} {\bibinfo {author} {\bibfnamefont {X.-G.}\ \bibnamefont
  {Wen}},\ }\href {\doibase 10.1103/PhysRevB.65.165113} {\bibfield  {journal}
  {\bibinfo  {journal} {Phys. Rev. B}\ }\textbf {\bibinfo {volume} {65}},\
  \bibinfo {pages} {165113} (\bibinfo {year} {2002}{\natexlab{a}})}\BibitemShut
  {NoStop}%
\bibitem [{\citenamefont {Wen}(2002{\natexlab{b}})}]{Wen2002PLA}%
  \BibitemOpen
  \bibfield  {author} {\bibinfo {author} {\bibfnamefont {X.-G.}\ \bibnamefont
  {Wen}},\ }\href {\doibase https://doi.org/10.1016/S0375-9601(02)00808-3}
  {\bibfield  {journal} {\bibinfo  {journal} {Phys. Lett. A}\ }\textbf
  {\bibinfo {volume} {300}},\ \bibinfo {pages} {175 } (\bibinfo {year}
  {2002}{\natexlab{b}})}\BibitemShut {NoStop}%
\bibitem [{\citenamefont {Kane}\ and\ \citenamefont
  {Mele}(2005{\natexlab{a}})}]{Kane2005sep}%
  \BibitemOpen
  \bibfield  {author} {\bibinfo {author} {\bibfnamefont {C.~L.}\ \bibnamefont
  {Kane}}\ and\ \bibinfo {author} {\bibfnamefont {E.~J.}\ \bibnamefont
  {Mele}},\ }\href {\doibase 10.1103/PhysRevLett.95.146802} {\bibfield
  {journal} {\bibinfo  {journal} {Phys. Rev. Lett.}\ }\textbf {\bibinfo
  {volume} {95}},\ \bibinfo {pages} {146802} (\bibinfo {year}
  {2005}{\natexlab{a}})}\BibitemShut {NoStop}%
\bibitem [{\citenamefont {Kane}\ and\ \citenamefont
  {Mele}(2005{\natexlab{b}})}]{Kane2005nov}%
  \BibitemOpen
  \bibfield  {author} {\bibinfo {author} {\bibfnamefont {C.~L.}\ \bibnamefont
  {Kane}}\ and\ \bibinfo {author} {\bibfnamefont {E.~J.}\ \bibnamefont
  {Mele}},\ }\href {\doibase 10.1103/PhysRevLett.95.226801} {\bibfield
  {journal} {\bibinfo  {journal} {Phys. Rev. Lett.}\ }\textbf {\bibinfo
  {volume} {95}},\ \bibinfo {pages} {226801} (\bibinfo {year}
  {2005}{\natexlab{b}})}\BibitemShut {NoStop}%
\bibitem [{\citenamefont {Fu}\ \emph {et~al.}(2007)\citenamefont {Fu},
  \citenamefont {Kane},\ and\ \citenamefont {Mele}}]{Fu20073d}%
  \BibitemOpen
  \bibfield  {author} {\bibinfo {author} {\bibfnamefont {L.}~\bibnamefont
  {Fu}}, \bibinfo {author} {\bibfnamefont {C.~L.}\ \bibnamefont {Kane}}, \ and\
  \bibinfo {author} {\bibfnamefont {E.~J.}\ \bibnamefont {Mele}},\ }\href
  {\doibase 10.1103/PhysRevLett.98.106803} {\bibfield  {journal} {\bibinfo
  {journal} {Phys. Rev. Lett.}\ }\textbf {\bibinfo {volume} {98}},\ \bibinfo
  {pages} {106803} (\bibinfo {year} {2007})}\BibitemShut {NoStop}%
\bibitem [{\citenamefont {Fu}\ and\ \citenamefont {Kane}(2007)}]{Fu2007inv}%
  \BibitemOpen
  \bibfield  {author} {\bibinfo {author} {\bibfnamefont {L.}~\bibnamefont
  {Fu}}\ and\ \bibinfo {author} {\bibfnamefont {C.~L.}\ \bibnamefont {Kane}},\
  }\href {\doibase 10.1103/PhysRevB.76.045302} {\bibfield  {journal} {\bibinfo
  {journal} {Phys. Rev. B}\ }\textbf {\bibinfo {volume} {76}},\ \bibinfo
  {pages} {045302} (\bibinfo {year} {2007})}\BibitemShut {NoStop}%
\bibitem [{\citenamefont {Yamada}\ \emph
  {et~al.}(2017{\natexlab{a}})\citenamefont {Yamada}, \citenamefont {Dwivedi},\
  and\ \citenamefont {Hermanns}}]{Yamada2017xsl}%
  \BibitemOpen
  \bibfield  {author} {\bibinfo {author} {\bibfnamefont {M.~G.}\ \bibnamefont
  {Yamada}}, \bibinfo {author} {\bibfnamefont {V.}~\bibnamefont {Dwivedi}}, \
  and\ \bibinfo {author} {\bibfnamefont {M.}~\bibnamefont {Hermanns}},\ }\href
  {\doibase 10.1103/PhysRevB.96.155107} {\bibfield  {journal} {\bibinfo
  {journal} {Phys. Rev. B}\ }\textbf {\bibinfo {volume} {96}},\ \bibinfo
  {pages} {155107} (\bibinfo {year} {2017}{\natexlab{a}})}\BibitemShut
  {NoStop}%
\bibitem [{\citenamefont {Qi}\ \emph {et~al.}()\citenamefont {Qi},
  \citenamefont {Cheng},\ and\ \citenamefont {Fang}}]{Qi2015}%
  \BibitemOpen
  \bibfield  {author} {\bibinfo {author} {\bibfnamefont {Y.}~\bibnamefont
  {Qi}}, \bibinfo {author} {\bibfnamefont {M.}~\bibnamefont {Cheng}}, \ and\
  \bibinfo {author} {\bibfnamefont {C.}~\bibnamefont {Fang}},\ }\href@noop {}
  {\ }\Eprint {http://arxiv.org/abs/1509.02927} {arXiv:1509.02927} \BibitemShut
  {NoStop}%
\bibitem [{\citenamefont {Lu}\ \emph {et~al.}(2017)\citenamefont {Lu},
  \citenamefont {Cho},\ and\ \citenamefont {Vishwanath}}]{Lu2017}%
  \BibitemOpen
  \bibfield  {author} {\bibinfo {author} {\bibfnamefont {Y.-M.}\ \bibnamefont
  {Lu}}, \bibinfo {author} {\bibfnamefont {G.~Y.}\ \bibnamefont {Cho}}, \ and\
  \bibinfo {author} {\bibfnamefont {A.}~\bibnamefont {Vishwanath}},\ }\href
  {\doibase 10.1103/PhysRevB.96.205150} {\bibfield  {journal} {\bibinfo
  {journal} {Phys. Rev. B}\ }\textbf {\bibinfo {volume} {96}},\ \bibinfo
  {pages} {205150} (\bibinfo {year} {2017})}\BibitemShut {NoStop}%
\bibitem [{\citenamefont {Essin}\ and\ \citenamefont
  {Hermele}(2013)}]{Essin2013}%
  \BibitemOpen
  \bibfield  {author} {\bibinfo {author} {\bibfnamefont {A.~M.}\ \bibnamefont
  {Essin}}\ and\ \bibinfo {author} {\bibfnamefont {M.}~\bibnamefont
  {Hermele}},\ }\href {\doibase 10.1103/PhysRevB.87.104406} {\bibfield
  {journal} {\bibinfo  {journal} {Phys. Rev. B}\ }\textbf {\bibinfo {volume}
  {87}},\ \bibinfo {pages} {104406} (\bibinfo {year} {2013})}\BibitemShut
  {NoStop}%
\bibitem [{\citenamefont {Kitaev}(2006)}]{Kitaev2006}%
  \BibitemOpen
  \bibfield  {author} {\bibinfo {author} {\bibfnamefont {A.}~\bibnamefont
  {Kitaev}},\ }\href {\doibase 10.1016/j.aop.2005.10.005} {\bibfield  {journal}
  {\bibinfo  {journal} {Ann. Phys.}\ }\textbf {\bibinfo {volume} {321}},\
  \bibinfo {pages} {2} (\bibinfo {year} {2006})},\ \bibinfo {note} {january
  Special Issue}\BibitemShut {NoStop}%
\bibitem [{\citenamefont {Kitaev}(2009)}]{Kitaev2009}%
  \BibitemOpen
  \bibfield  {author} {\bibinfo {author} {\bibfnamefont {A.}~\bibnamefont
  {Kitaev}},\ }\href {\doibase 10.1063/1.3149495} {\bibfield  {journal}
  {\bibinfo  {journal} {AIP Conf. Proc.}\ }\textbf {\bibinfo {volume} {1134}},\
  \bibinfo {pages} {22} (\bibinfo {year} {2009})}\BibitemShut {NoStop}%
\bibitem [{\citenamefont {Ryu}\ \emph {et~al.}(2010)\citenamefont {Ryu},
  \citenamefont {Schnyder}, \citenamefont {Furusaki},\ and\ \citenamefont
  {Ludwig}}]{Ryu2010}%
  \BibitemOpen
  \bibfield  {author} {\bibinfo {author} {\bibfnamefont {S.}~\bibnamefont
  {Ryu}}, \bibinfo {author} {\bibfnamefont {A.~P.}\ \bibnamefont {Schnyder}},
  \bibinfo {author} {\bibfnamefont {A.}~\bibnamefont {Furusaki}}, \ and\
  \bibinfo {author} {\bibfnamefont {A.~W.~W.}\ \bibnamefont {Ludwig}},\ }\href
  {\doibase 10.1088/1367-2630/12/6/065010} {\bibfield  {journal} {\bibinfo
  {journal} {New J. Phys.}\ }\textbf {\bibinfo {volume} {12}},\ \bibinfo
  {pages} {065010} (\bibinfo {year} {2010})}\BibitemShut {NoStop}%
\bibitem [{\citenamefont {Yang}\ \emph {et~al.}(2007)\citenamefont {Yang},
  \citenamefont {Zhou},\ and\ \citenamefont {Sun}}]{Yang2007}%
  \BibitemOpen
  \bibfield  {author} {\bibinfo {author} {\bibfnamefont {S.}~\bibnamefont
  {Yang}}, \bibinfo {author} {\bibfnamefont {D.~L.}\ \bibnamefont {Zhou}}, \
  and\ \bibinfo {author} {\bibfnamefont {C.~P.}\ \bibnamefont {Sun}},\ }\href
  {\doibase 10.1103/PhysRevB.76.180404} {\bibfield  {journal} {\bibinfo
  {journal} {Phys. Rev. B}\ }\textbf {\bibinfo {volume} {76}},\ \bibinfo
  {pages} {180404} (\bibinfo {year} {2007})}\BibitemShut {NoStop}%
\bibitem [{\citenamefont {Baskaran}\ \emph {et~al.}()\citenamefont {Baskaran},
  \citenamefont {Santhosh},\ and\ \citenamefont {Shankar}}]{Baskaran2009}%
  \BibitemOpen
  \bibfield  {author} {\bibinfo {author} {\bibfnamefont {G.}~\bibnamefont
  {Baskaran}}, \bibinfo {author} {\bibfnamefont {G.}~\bibnamefont {Santhosh}},
  \ and\ \bibinfo {author} {\bibfnamefont {R.}~\bibnamefont {Shankar}},\
  }\href@noop {} {\ }\Eprint {http://arxiv.org/abs/0908.1614} {arXiv:0908.1614}
  \BibitemShut {NoStop}%
\bibitem [{\citenamefont {Kells}\ \emph {et~al.}(2011)\citenamefont {Kells},
  \citenamefont {Kailasvuori}, \citenamefont {Slingerland},\ and\ \citenamefont
  {Vala}}]{Kells2011}%
  \BibitemOpen
  \bibfield  {author} {\bibinfo {author} {\bibfnamefont {G.}~\bibnamefont
  {Kells}}, \bibinfo {author} {\bibfnamefont {J.}~\bibnamefont {Kailasvuori}},
  \bibinfo {author} {\bibfnamefont {J.~K.}\ \bibnamefont {Slingerland}}, \ and\
  \bibinfo {author} {\bibfnamefont {J.}~\bibnamefont {Vala}},\ }\href {\doibase
  10.1088/1367-2630/13/9/095014} {\bibfield  {journal} {\bibinfo  {journal}
  {New J. Phys.}\ }\textbf {\bibinfo {volume} {13}},\ \bibinfo {pages} {095014}
  (\bibinfo {year} {2011})}\BibitemShut {NoStop}%
\bibitem [{\citenamefont {Hermanns}\ and\ \citenamefont
  {Trebst}(2014)}]{Hermanns2014}%
  \BibitemOpen
  \bibfield  {author} {\bibinfo {author} {\bibfnamefont {M.}~\bibnamefont
  {Hermanns}}\ and\ \bibinfo {author} {\bibfnamefont {S.}~\bibnamefont
  {Trebst}},\ }\href {\doibase 10.1103/PhysRevB.89.235102} {\bibfield
  {journal} {\bibinfo  {journal} {Phys. Rev. B}\ }\textbf {\bibinfo {volume}
  {89}},\ \bibinfo {pages} {235102} (\bibinfo {year} {2014})}\BibitemShut
  {NoStop}%
\bibitem [{\citenamefont {Hermanns}\ \emph {et~al.}(2015)\citenamefont
  {Hermanns}, \citenamefont {Trebst},\ and\ \citenamefont
  {Rosch}}]{Hermanns2015}%
  \BibitemOpen
  \bibfield  {author} {\bibinfo {author} {\bibfnamefont {M.}~\bibnamefont
  {Hermanns}}, \bibinfo {author} {\bibfnamefont {S.}~\bibnamefont {Trebst}}, \
  and\ \bibinfo {author} {\bibfnamefont {A.}~\bibnamefont {Rosch}},\ }\href
  {\doibase 10.1103/PhysRevLett.115.177205} {\bibfield  {journal} {\bibinfo
  {journal} {Phys. Rev. Lett.}\ }\textbf {\bibinfo {volume} {115}},\ \bibinfo
  {pages} {177205} (\bibinfo {year} {2015})}\BibitemShut {NoStop}%
\bibitem [{\citenamefont {Lieb}(1994)}]{Lieb1994}%
  \BibitemOpen
  \bibfield  {author} {\bibinfo {author} {\bibfnamefont {E.~H.}\ \bibnamefont
  {Lieb}},\ }\href {\doibase 10.1103/PhysRevLett.73.2158} {\bibfield  {journal}
  {\bibinfo  {journal} {Phys. Rev. Lett.}\ }\textbf {\bibinfo {volume} {73}},\
  \bibinfo {pages} {2158} (\bibinfo {year} {1994})}\BibitemShut {NoStop}%
\bibitem [{\citenamefont {O'Brien}\ \emph {et~al.}(2016)\citenamefont
  {O'Brien}, \citenamefont {Hermanns},\ and\ \citenamefont
  {Trebst}}]{Obrien2016}%
  \BibitemOpen
  \bibfield  {author} {\bibinfo {author} {\bibfnamefont {K.}~\bibnamefont
  {O'Brien}}, \bibinfo {author} {\bibfnamefont {M.}~\bibnamefont {Hermanns}}, \
  and\ \bibinfo {author} {\bibfnamefont {S.}~\bibnamefont {Trebst}},\ }\href
  {\doibase 10.1103/PhysRevB.93.085101} {\bibfield  {journal} {\bibinfo
  {journal} {Phys. Rev. B}\ }\textbf {\bibinfo {volume} {93}},\ \bibinfo
  {pages} {085101} (\bibinfo {year} {2016})}\BibitemShut {NoStop}%
\bibitem [{\citenamefont {Fu}(2011)}]{Fu2011}%
  \BibitemOpen
  \bibfield  {author} {\bibinfo {author} {\bibfnamefont {L.}~\bibnamefont
  {Fu}},\ }\href {\doibase 10.1103/PhysRevLett.106.106802} {\bibfield
  {journal} {\bibinfo  {journal} {Phys. Rev. Lett.}\ }\textbf {\bibinfo
  {volume} {106}},\ \bibinfo {pages} {106802} (\bibinfo {year}
  {2011})}\BibitemShut {NoStop}%
\bibitem [{Note1()}]{Note1}%
  \BibitemOpen
  \bibinfo {note} {See Supplemental Material at [URL will be inserted by
  publisher] for Fig.~S1 and the PSG classification.}\BibitemShut {Stop}%
\bibitem [{\citenamefont {Watanabe}\ \emph {et~al.}(2018)\citenamefont
  {Watanabe}, \citenamefont {Po},\ and\ \citenamefont
  {Vishwanath}}]{Watanabe2018}%
  \BibitemOpen
  \bibfield  {author} {\bibinfo {author} {\bibfnamefont {H.}~\bibnamefont
  {Watanabe}}, \bibinfo {author} {\bibfnamefont {H.~C.}\ \bibnamefont {Po}}, \
  and\ \bibinfo {author} {\bibfnamefont {A.}~\bibnamefont {Vishwanath}},\
  }\href {\doibase 10.1126/sciadv.aat8685} {\bibfield  {journal} {\bibinfo
  {journal} {Sci. Adv.}\ }\textbf {\bibinfo {volume} {4}} (\bibinfo {year}
  {2018}),\ 10.1126/sciadv.aat8685}\BibitemShut {NoStop}%
\bibitem [{\citenamefont {Mong}\ \emph {et~al.}(2010)\citenamefont {Mong},
  \citenamefont {Essin},\ and\ \citenamefont {Moore}}]{Mong2010}%
  \BibitemOpen
  \bibfield  {author} {\bibinfo {author} {\bibfnamefont {R.~S.~K.}\
  \bibnamefont {Mong}}, \bibinfo {author} {\bibfnamefont {A.~M.}\ \bibnamefont
  {Essin}}, \ and\ \bibinfo {author} {\bibfnamefont {J.~E.}\ \bibnamefont
  {Moore}},\ }\href {\doibase 10.1103/PhysRevB.81.245209} {\bibfield  {journal}
  {\bibinfo  {journal} {Phys. Rev. B}\ }\textbf {\bibinfo {volume} {81}},\
  \bibinfo {pages} {245209} (\bibinfo {year} {2010})}\BibitemShut {NoStop}%
\bibitem [{\citenamefont {Fang}\ \emph {et~al.}(2013)\citenamefont {Fang},
  \citenamefont {Gilbert},\ and\ \citenamefont {Bernevig}}]{Fang2013}%
  \BibitemOpen
  \bibfield  {author} {\bibinfo {author} {\bibfnamefont {C.}~\bibnamefont
  {Fang}}, \bibinfo {author} {\bibfnamefont {M.~J.}\ \bibnamefont {Gilbert}}, \
  and\ \bibinfo {author} {\bibfnamefont {B.~A.}\ \bibnamefont {Bernevig}},\
  }\href {\doibase 10.1103/PhysRevB.88.085406} {\bibfield  {journal} {\bibinfo
  {journal} {Phys. Rev. B}\ }\textbf {\bibinfo {volume} {88}},\ \bibinfo
  {pages} {085406} (\bibinfo {year} {2013})}\BibitemShut {NoStop}%
\bibitem [{\citenamefont {Budich}\ and\ \citenamefont
  {Ardonne}(2013)}]{Budich2013diii}%
  \BibitemOpen
  \bibfield  {author} {\bibinfo {author} {\bibfnamefont {J.~C.}\ \bibnamefont
  {Budich}}\ and\ \bibinfo {author} {\bibfnamefont {E.}~\bibnamefont
  {Ardonne}},\ }\href {\doibase 10.1103/PhysRevB.88.134523} {\bibfield
  {journal} {\bibinfo  {journal} {Phys. Rev. B}\ }\textbf {\bibinfo {volume}
  {88}},\ \bibinfo {pages} {134523} (\bibinfo {year} {2013})}\BibitemShut
  {NoStop}%
\bibitem [{\citenamefont {Wang}\ \emph {et~al.}()\citenamefont {Wang},
  \citenamefont {Fang}, \citenamefont {Cheng}, \citenamefont {Qi},\ and\
  \citenamefont {Meng}}]{Wang2017}%
  \BibitemOpen
  \bibfield  {author} {\bibinfo {author} {\bibfnamefont {Y.-C.}\ \bibnamefont
  {Wang}}, \bibinfo {author} {\bibfnamefont {C.}~\bibnamefont {Fang}}, \bibinfo
  {author} {\bibfnamefont {M.}~\bibnamefont {Cheng}}, \bibinfo {author}
  {\bibfnamefont {Y.}~\bibnamefont {Qi}}, \ and\ \bibinfo {author}
  {\bibfnamefont {Z.~Y.}\ \bibnamefont {Meng}},\ }\href@noop {} {\ }\Eprint
  {http://arxiv.org/abs/1701.01552} {arXiv:1701.01552} \BibitemShut {NoStop}%
\bibitem [{\citenamefont {Shiozaki}\ \emph {et~al.}(2016)\citenamefont
  {Shiozaki}, \citenamefont {Sato},\ and\ \citenamefont {Gomi}}]{Shiozaki2016}%
  \BibitemOpen
  \bibfield  {author} {\bibinfo {author} {\bibfnamefont {K.}~\bibnamefont
  {Shiozaki}}, \bibinfo {author} {\bibfnamefont {M.}~\bibnamefont {Sato}}, \
  and\ \bibinfo {author} {\bibfnamefont {K.}~\bibnamefont {Gomi}},\ }\href
  {\doibase 10.1103/PhysRevB.93.195413} {\bibfield  {journal} {\bibinfo
  {journal} {Phys. Rev. B}\ }\textbf {\bibinfo {volume} {93}},\ \bibinfo
  {pages} {195413} (\bibinfo {year} {2016})}\BibitemShut {NoStop}%
\bibitem [{\citenamefont {Morimoto}\ \emph {et~al.}(2015)\citenamefont
  {Morimoto}, \citenamefont {Furusaki},\ and\ \citenamefont
  {Mudry}}]{Morimoto2015}%
  \BibitemOpen
  \bibfield  {author} {\bibinfo {author} {\bibfnamefont {T.}~\bibnamefont
  {Morimoto}}, \bibinfo {author} {\bibfnamefont {A.}~\bibnamefont {Furusaki}},
  \ and\ \bibinfo {author} {\bibfnamefont {C.}~\bibnamefont {Mudry}},\ }\href
  {\doibase 10.1103/PhysRevB.91.235111} {\bibfield  {journal} {\bibinfo
  {journal} {Phys. Rev. B}\ }\textbf {\bibinfo {volume} {91}},\ \bibinfo
  {pages} {235111} (\bibinfo {year} {2015})}\BibitemShut {NoStop}%
\bibitem [{\citenamefont {Yamada}(2020)}]{Yamada2020}%
  \BibitemOpen
  \bibfield  {author} {\bibinfo {author} {\bibfnamefont {M.~G.}\ \bibnamefont
  {Yamada}},\ }\href {\doibase 10.1038/s41535-020-00285-3} {\bibfield
  {journal} {\bibinfo  {journal} {npj Quantum Mater.}\ }\textbf {\bibinfo
  {volume} {5}},\ \bibinfo {pages} {82} (\bibinfo {year} {2020})}\BibitemShut
  {NoStop}%
\bibitem [{\citenamefont {Yamada}\ \emph {et~al.}(2018)\citenamefont {Yamada},
  \citenamefont {Oshikawa},\ and\ \citenamefont {Jackeli}}]{Yamada2018}%
  \BibitemOpen
  \bibfield  {author} {\bibinfo {author} {\bibfnamefont {M.~G.}\ \bibnamefont
  {Yamada}}, \bibinfo {author} {\bibfnamefont {M.}~\bibnamefont {Oshikawa}}, \
  and\ \bibinfo {author} {\bibfnamefont {G.}~\bibnamefont {Jackeli}},\ }\href
  {\doibase 10.1103/PhysRevLett.121.097201} {\bibfield  {journal} {\bibinfo
  {journal} {Phys. Rev. Lett.}\ }\textbf {\bibinfo {volume} {121}},\ \bibinfo
  {pages} {097201} (\bibinfo {year} {2018})}\BibitemShut {NoStop}%
\bibitem [{\citenamefont {Dwivedi}\ \emph {et~al.}(2018)\citenamefont
  {Dwivedi}, \citenamefont {Hickey}, \citenamefont {Eschmann},\ and\
  \citenamefont {Trebst}}]{Dwivedi2018}%
  \BibitemOpen
  \bibfield  {author} {\bibinfo {author} {\bibfnamefont {V.}~\bibnamefont
  {Dwivedi}}, \bibinfo {author} {\bibfnamefont {C.}~\bibnamefont {Hickey}},
  \bibinfo {author} {\bibfnamefont {T.}~\bibnamefont {Eschmann}}, \ and\
  \bibinfo {author} {\bibfnamefont {S.}~\bibnamefont {Trebst}},\ }\href
  {\doibase 10.1103/PhysRevB.98.054432} {\bibfield  {journal} {\bibinfo
  {journal} {Phys. Rev. B}\ }\textbf {\bibinfo {volume} {98}},\ \bibinfo
  {pages} {054432} (\bibinfo {year} {2018})}\BibitemShut {NoStop}%
\bibitem [{\citenamefont {Zhao}\ \emph {et~al.}()\citenamefont {Zhao},
  \citenamefont {Lu},\ and\ \citenamefont {Yang}}]{Zhao2020}%
  \BibitemOpen
  \bibfield  {author} {\bibinfo {author} {\bibfnamefont {Y.~X.}\ \bibnamefont
  {Zhao}}, \bibinfo {author} {\bibfnamefont {Y.}~\bibnamefont {Lu}}, \ and\
  \bibinfo {author} {\bibfnamefont {S.~A.}\ \bibnamefont {Yang}},\ }\href@noop
  {} {\ }\Eprint {http://arxiv.org/abs/2005.14500} {arXiv:2005.14500}
  \BibitemShut {NoStop}%
\bibitem [{\citenamefont {Jackeli}\ and\ \citenamefont
  {Khaliullin}(2009)}]{Jackeli2009}%
  \BibitemOpen
  \bibfield  {author} {\bibinfo {author} {\bibfnamefont {G.}~\bibnamefont
  {Jackeli}}\ and\ \bibinfo {author} {\bibfnamefont {G.}~\bibnamefont
  {Khaliullin}},\ }\href {\doibase 10.1103/PhysRevLett.102.017205} {\bibfield
  {journal} {\bibinfo  {journal} {Phys. Rev. Lett.}\ }\textbf {\bibinfo
  {volume} {102}},\ \bibinfo {pages} {017205} (\bibinfo {year}
  {2009})}\BibitemShut {NoStop}%
\bibitem [{\citenamefont {Plumb}\ \emph {et~al.}(2014)\citenamefont {Plumb},
  \citenamefont {Clancy}, \citenamefont {Sandilands}, \citenamefont {Shankar},
  \citenamefont {Hu}, \citenamefont {Burch}, \citenamefont {Kee},\ and\
  \citenamefont {Kim}}]{Plumb2014}%
  \BibitemOpen
  \bibfield  {author} {\bibinfo {author} {\bibfnamefont {K.~W.}\ \bibnamefont
  {Plumb}}, \bibinfo {author} {\bibfnamefont {J.~P.}\ \bibnamefont {Clancy}},
  \bibinfo {author} {\bibfnamefont {L.~J.}\ \bibnamefont {Sandilands}},
  \bibinfo {author} {\bibfnamefont {V.~V.}\ \bibnamefont {Shankar}}, \bibinfo
  {author} {\bibfnamefont {Y.~F.}\ \bibnamefont {Hu}}, \bibinfo {author}
  {\bibfnamefont {K.~S.}\ \bibnamefont {Burch}}, \bibinfo {author}
  {\bibfnamefont {H.-Y.}\ \bibnamefont {Kee}}, \ and\ \bibinfo {author}
  {\bibfnamefont {Y.-J.}\ \bibnamefont {Kim}},\ }\href {\doibase
  10.1103/PhysRevB.90.041112} {\bibfield  {journal} {\bibinfo  {journal} {Phys.
  Rev. B}\ }\textbf {\bibinfo {volume} {90}},\ \bibinfo {pages} {041112}
  (\bibinfo {year} {2014})}\BibitemShut {NoStop}%
\bibitem [{\citenamefont {Yamada}\ \emph
  {et~al.}(2017{\natexlab{b}})\citenamefont {Yamada}, \citenamefont {Fujita},\
  and\ \citenamefont {Oshikawa}}]{Yamada2017mof}%
  \BibitemOpen
  \bibfield  {author} {\bibinfo {author} {\bibfnamefont {M.~G.}\ \bibnamefont
  {Yamada}}, \bibinfo {author} {\bibfnamefont {H.}~\bibnamefont {Fujita}}, \
  and\ \bibinfo {author} {\bibfnamefont {M.}~\bibnamefont {Oshikawa}},\ }\href
  {\doibase 10.1103/PhysRevLett.119.057202} {\bibfield  {journal} {\bibinfo
  {journal} {Phys. Rev. Lett.}\ }\textbf {\bibinfo {volume} {119}},\ \bibinfo
  {pages} {057202} (\bibinfo {year} {2017}{\natexlab{b}})}\BibitemShut
  {NoStop}%
\end{thebibliography}%


\begin{thebibliography}{6}%
\makeatletter
\providecommand \@ifxundefined [1]{%
 \@ifx{#1\undefined}
}%
\providecommand \@ifnum [1]{%
 \ifnum #1\expandafter \@firstoftwo
 \else \expandafter \@secondoftwo
 \fi
}%
\providecommand \@ifx [1]{%
 \ifx #1\expandafter \@firstoftwo
 \else \expandafter \@secondoftwo
 \fi
}%
\providecommand \natexlab [1]{#1}%
\providecommand \enquote  [1]{``#1''}%
\providecommand \bibnamefont  [1]{#1}%
\providecommand \bibfnamefont [1]{#1}%
\providecommand \citenamefont [1]{#1}%
\providecommand \href@noop [0]{\@secondoftwo}%
\providecommand \href [0]{\begingroup \@sanitize@url \@href}%
\providecommand \@href[1]{\@@startlink{#1}\@@href}%
\providecommand \@@href[1]{\endgroup#1\@@endlink}%
\providecommand \@sanitize@url [0]{\catcode `\\12\catcode `\$12\catcode
  `\&12\catcode `\#12\catcode `\^12\catcode `\_12\catcode `\%12\relax}%
\providecommand \@@startlink[1]{}%
\providecommand \@@endlink[0]{}%
\providecommand \url  [0]{\begingroup\@sanitize@url \@url }%
\providecommand \@url [1]{\endgroup\@href {#1}{\urlprefix }}%
\providecommand \urlprefix  [0]{URL }%
\providecommand \Eprint [0]{\href }%
\providecommand \doibase [0]{http://dx.doi.org/}%
\providecommand \selectlanguage [0]{\@gobble}%
\providecommand \bibinfo  [0]{\@secondoftwo}%
\providecommand \bibfield  [0]{\@secondoftwo}%
\providecommand \translation [1]{[#1]}%
\providecommand \BibitemOpen [0]{}%
\providecommand \bibitemStop [0]{}%
\providecommand \bibitemNoStop [0]{.\EOS\space}%
\providecommand \EOS [0]{\spacefactor3000\relax}%
\providecommand \BibitemShut  [1]{\csname bibitem#1\endcsname}%
\let\auto@bib@innerbib\@empty
\bibitem [{\citenamefont {You}\ \emph {et~al.}(2012)\citenamefont {You},
  \citenamefont {Kimchi},\ and\ \citenamefont {Vishwanath}}]{You2012}%
  \BibitemOpen
  \bibfield  {author} {\bibinfo {author} {\bibfnamefont {Y.-Z.}\ \bibnamefont
  {You}}, \bibinfo {author} {\bibfnamefont {I.}~\bibnamefont {Kimchi}}, \ and\
  \bibinfo {author} {\bibfnamefont {A.}~\bibnamefont {Vishwanath}},\ }\href
  {\doibase 10.1103/PhysRevB.86.085145} {\bibfield  {journal} {\bibinfo
  {journal} {Phys. Rev. B}\ }\textbf {\bibinfo {volume} {86}},\ \bibinfo
  {pages} {085145} (\bibinfo {year} {2012})}\BibitemShut {NoStop}%
\bibitem [{\citenamefont {Seifert}\ \emph {et~al.}(2018)\citenamefont
  {Seifert}, \citenamefont {Meng},\ and\ \citenamefont {Vojta}}]{Seifert2018}%
  \BibitemOpen
  \bibfield  {author} {\bibinfo {author} {\bibfnamefont {U.~F.~P.}\
  \bibnamefont {Seifert}}, \bibinfo {author} {\bibfnamefont {T.}~\bibnamefont
  {Meng}}, \ and\ \bibinfo {author} {\bibfnamefont {M.}~\bibnamefont {Vojta}},\
  }\href {\doibase 10.1103/PhysRevB.97.085118} {\bibfield  {journal} {\bibinfo
  {journal} {Phys. Rev. B}\ }\textbf {\bibinfo {volume} {97}},\ \bibinfo
  {pages} {085118} (\bibinfo {year} {2018})}\BibitemShut {NoStop}%
\bibitem [{\citenamefont {O'Brien}(2019)}]{Obrien2019}%
  \BibitemOpen
  \bibfield  {author} {\bibinfo {author} {\bibfnamefont {K.~M.}\ \bibnamefont
  {O'Brien}},\ }\href@noop {} {\enquote {\bibinfo {title}
  {{Three}-{Dimensional} {Kitaev} {Spin} {Liquids}},}\ } (\bibinfo {year}
  {2019}),\ \bibinfo {note} {the University of Cologne, PhD Thesis}\BibitemShut
  {NoStop}%
\bibitem [{\citenamefont {Wen}(2002)}]{Wen2002PRB}%
  \BibitemOpen
  \bibfield  {author} {\bibinfo {author} {\bibfnamefont {X.-G.}\ \bibnamefont
  {Wen}},\ }\href {\doibase 10.1103/PhysRevB.65.165113} {\bibfield  {journal}
  {\bibinfo  {journal} {Phys. Rev. B}\ }\textbf {\bibinfo {volume} {65}},\
  \bibinfo {pages} {165113} (\bibinfo {year} {2002})}\BibitemShut {NoStop}%
\bibitem [{\citenamefont {Lieb}(1994)}]{Lieb1994}%
  \BibitemOpen
  \bibfield  {author} {\bibinfo {author} {\bibfnamefont {E.~H.}\ \bibnamefont
  {Lieb}},\ }\href {\doibase 10.1103/PhysRevLett.73.2158} {\bibfield  {journal}
  {\bibinfo  {journal} {Phys. Rev. Lett.}\ }\textbf {\bibinfo {volume} {73}},\
  \bibinfo {pages} {2158} (\bibinfo {year} {1994})}\BibitemShut {NoStop}%
\bibitem [{\citenamefont {Fu}\ and\ \citenamefont {Kane}(2007)}]{Fu2007inv}%
  \BibitemOpen
  \bibfield  {author} {\bibinfo {author} {\bibfnamefont {L.}~\bibnamefont
  {Fu}}\ and\ \bibinfo {author} {\bibfnamefont {C.~L.}\ \bibnamefont {Kane}},\
  }\href {\doibase 10.1103/PhysRevB.76.045302} {\bibfield  {journal} {\bibinfo
  {journal} {Phys. Rev. B}\ }\textbf {\bibinfo {volume} {76}},\ \bibinfo
  {pages} {045302} (\bibinfo {year} {2007})}\BibitemShut {NoStop}%
\end{thebibliography}%

\end{document}


\title{Supplemental Material for ``Topological $Z_2$ invariant in Kitaev spin liquids: \\
Classification of gapped spin liquids beyond projective symmetry group''}

\author{Masahiko G. Yamada}
\affiliation{Department of Materials Engineering Science, Osaka University, Toyonaka 560-8531, Japan}
\affiliation{Institute for Solid State Physics, University of Tokyo, Kashiwa 277-8581, Japan}

\maketitle

\section{Projective symmetry group}

\begin{figure}
\centering
\includegraphics[width=4cm]{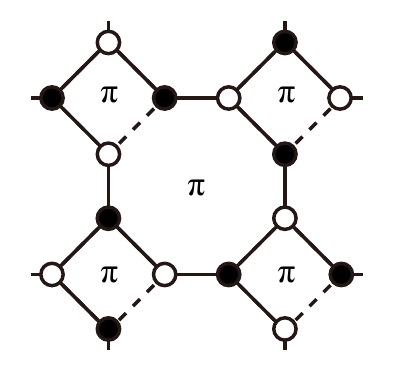}
\caption{Squareoctagon lattice with a $\pi$ flux.  Sublattice A is shown in white
circles and sublattice B is shown in black circles.  Solid bonds have $+1$ hopping,
while dashed bonds have $-1$ hopping to make the flux value $\pi$.}
\label{piflux}
\end{figure}

We present the projective symmetry group (PSG) of the Kitaev model on the squareoctagon
lattice, assuming the $\pi$-flux mean-field theory~\cite{You2012,Seifert2018,Obrien2019}.
We do not include other flux sectors because they are not discussed in the main text.
We also do not follow Wen's approach to the Heisenberg model~\cite{Wen2002PRB}.
We would not exhaust
all possible PSGs in the Kitaev model because in the Kitaev model the exact solution
is only one determined by Lieb's theorem~\cite{Lieb1994}.  All the other mean-field
solutions are guaranteed to have a higher energy and of less importance.

First let us introduce four Majorana
operators $c_j^\gamma$ for the $j$th site with $\gamma = 0,\,x,\,y,\,z$.
$c_j^0$ corresponds to an itinerant Majorana fermion $c_j$ in the main text,
so we sometimes omit the superscript 0 for simplicity.  After introducing
Kitaev's representation $\sigma^\gamma=ic^\gamma c^0$, the Kitaev Hamiltonian
is recast into a four-body Majorana Hamiltonian.  Its mean-field solution has a form
\begin{equation}
    H_\textrm{MF} = -\frac{1}{8} \sum_{\langle jk \rangle \in \gamma} J_\gamma
    \left( ic_j \overline{U}_{jk}^{\gamma \gamma} c_k +
    ic_j^\gamma \overline{U}_{jk}^{0 0} c_k^\gamma \right) + \textrm{Const}.,
\end{equation}
where $\overline{U}_{jk}^{\mu\nu} = \langle ic_j^\mu c_k^\nu \rangle$.

At the center of the phase diagram $J_x=J_y=J_z>0$, the solution with a $\pi$ flux is
\begin{equation}
    \overline{U}_{jk}^{\mu\nu}=
    \begin{cases}
        \textrm{Const.} & \textrm{for}\;\mu=\nu=0 \\
        \pm 1 & \textrm{for}\;\mu=\nu=\gamma\;\textrm{and}\;\langle jk \rangle \in \gamma \\
        0 & \textrm{otherwise}
    \end{cases},
\end{equation}
where the sign of $\pm 1$ is decided as $+1$ (resp. $-1$) for a solid (resp. dashed) bond
in Fig.~\ref{piflux}, $j$ always belongs to sublattice A and $k$ belongs to sublattice B, and
$\overline{U}_{jk}^{\mu\nu} = -\overline{U}_{kj}^{\nu\mu}$~\cite{Obrien2019}.
In the generic case, $|\overline{U}_{jk}^{\mu\nu}|$ will depend on each ``bond color'', but
this does not matter as long as the translation and time-reversal symmetries
are not broken and $\overline{U}_{jk}$ stays diagonal.

Assuming all of the relevant symmetries in the main text, \textit{i.e.} the translation
and time-reversal symmetries, we now have 3 symmetry group (SG) generators:
$SG = \langle \{ \hat{T}_1, \hat{T}_2, \hat{\Theta}\} \rangle$.
We note that $\hat{T}_1$ and $\hat{T}_2$ are translations along the $x$- and $y$-directions,
respectively, and $\hat{\Theta}$ is time reversal.
Here the meaning of the symmetry operation is slightly different from that in the main text,
so we put a hat to distinguish this mean-field theory from the exact solution.

The gauge chosen in Fig.~\ref{piflux} respects the translation symmetries,
which simplifies discussions drastically.  The translation along the $x$-direction
only changes the sublattice parity, so
\begin{equation}
    \hat{T}_1(i\overline{U}_{jk}) = i\overline{U}_{kj} = -i\overline{U}_{jk}.
\end{equation}
The same is true for the $y$-direction.
\begin{equation}
    \hat{T}_2(i\overline{U}_{jk}) = i\overline{U}_{kj} = -i\overline{U}_{jk}.
\end{equation}
Finally, since the time-reversal symmetry is antiunitary,
\begin{equation}
    \hat{\Theta}(i\overline{U}_{jk}) = -i\overline{U}_{jk}.
\end{equation}
As described in the main text, the gauge transformation accompanied by these
operations is always $G(j) = (-1)^j$ where $(-1)^j$ is a sublattice parity.
If we write this site-dependent gauge transformation as $\hat{G},$ then
the PSG for the Kitaev model on the $\pi$-flux squareoctagon lattice
with a symmetry group $SG$ is given by
\begin{equation}
    PSG = \langle \{ \hat{G}_0, \hat{G}\hat{T}_1, \hat{G}\hat{T}_2, \hat{G}\hat{\Theta} \} \rangle,
\end{equation}
where $G_0(j) = -1$ generates the invariant gauge group.

An important point is that the discussion here applies to any value of
$J_x>0$, $J_y>0$, and $J_z>0$.  Indeed, the $\pi$-flux ansatz is always guaranteed
to be the ground state by Lieb's theorem~\cite{Lieb1994}, and thus $PSG$ is kept constant
in the whole phase diagram of $J_x>0$, $J_y>0$, and $J_z>0$.  In the language of
PSG, there is no distinction between phases $A_1$ and $A_2$ in the main text.
They are different only in the sense of the symmetry-protected topological order
of fermionic spinon excitations, and therefore we claim it to be a classification
of gapped quantum spin liquids beyond PSG.

\section{Nonlocal Pfaffian invariant approach}

As mentioned in the main text, we would introduce the third derivation of the
$Z_2$ invariant, a nonlocal Pfaffian invariant approach.
This section is important in the sense that the Pfaffian invariant is modified
reflecting the nonlocality of the time-reversal action in the $k$-space.
We believe this phenomenon is intrinsic to quantum spin liquids with a nontrivial PSG
and thus is of great importance as a guiding principle to explore another
topological invariant of quantum spin liquids in the future.

We already found a Fu-Kane-type formula for the classification of a
class of Kitaev models in the main text, and thus we can expect that it is related to some
Pfaffian invariant~\cite{Fu2007inv}.  Differently from the topological insulator, a new
Pfaffian invariant requires a quantity defined nonlocally in the $k$-space.
Thus, we call it nonlocal Pfaffian invariant, defined from a vector bundle on the reduced
(half) Brillouin zone.  In this way we can directly connect a Fu-Kane invariant
to a Berry phase in this half Brillouin zone (HBZ) [see Fig.~2 in the main text].
\begin{equation}
    \mathcal{A}(\bm{k}) = -i \sum_{\alpha} \braket{u_{\alpha,\bm{k}}|\nabla_{\bm{k}}|u_{\alpha,\bm{k}}}.
\end{equation}

Although the time reversal itself divides the Brillouin zone into an orbifold by identifying
$\bm{k}$ and $\bm{k}_0 - \bm{k}$, the Brillouin zone is now divided into
a 2D torus by combining the time reversal and the inversion, which identifies
$\bm{k}$ and $\bm{k} + \bm{k}_0$.  Thus, on this ``half torus'', shown in the pink shaded
region in Fig.~2 in the main text, we can define a smooth twofold degenerate
vector bundle of eigenstates, \textit{i.e.} the Hilbert spaces for Bloch functions
$\mathcal{H}_{\bm{k}}$ and $\mathcal{H}_{\bm{k}+\bm{k}_0}$ are combined into
$\mathcal{H}_{\bm{k}} \oplus \mathcal{H}_{\bm{k}+\bm{k}_0}$.
Now we can identify two points $(k_x,\,k_y)$ and $(k_x + \pi,\,k_y + \pi)$
in the original Brillouin zone to make an HBZ, and we distinguish two positions
$(k_x,\,k_y)$ and $(k_x + \pi,\,k_y + \pi)$
by an internal degree of freedom $\tau = \;\uparrow,\,\downarrow$, respectively.
From now on $\tau$ is regarded as an internal pseudospin, but offdiagonal components
about $\tau$ is actually nonlocal in the original $k$-space.
On this manifold of HBZ, the time-reversal switches the internal degree of freedom,
so the time reversal in HBZ can be written $\Theta_+ = \tau^x \otimes \Theta$.

About the Bloch Hamiltonian $H_\textrm{HBZ}=H(\bm{k}) \oplus H(\bm{k}+\bm{k}_0)$,
$H_\textrm{HBZ}$ indeed commutes with $\Theta_+$ at IIM.  However, there are
another symmetry $\tau^z$, which commutes with the Bloch Hamiltonian in the whole HBZ.
Thus, we can define another time-reversal symmetry $\Theta_-$ by
$\Theta_- = \tau^z \Theta_+ = i\tau^y \otimes \Theta$.  Here $\Theta_-^2 = -1$.

The parity time-reversal ($\Theta_+$) operation $\Theta_+ P$ also acts
antiunitarily with a condition $(\Theta_+ P)^2 = -1$, and thus we can use
these operators to define a $Z_2$ invariant.
We define $v_{mn}(\bm{k})=\braket{u_{m,\bm{k}}|\Theta_+ P|u_{n,\bm{k}}}$.
Here we newly include $\tau$ indices in $m$ and $n.$  Since
$[\Theta_+ P, H_\textrm{HBZ}]=0$, a matrix $v(\bm{k})$ is unitary, and
from $(\Theta_+ P)^2 = -1$, $v(\bm{k})$ is antisymmetric.  Thus, the Pfaffian
of $v(\bm{k})$ exists and has a unit magnitude.  Its gradient should be related to
a Berry phase.
\begin{align}
    \mathcal{A}(\bm{k}) + \mathcal{A}(\bm{k}+\bm{k}_0) &= -\frac{i}{2} \textrm{Tr}\,[v(\bm{k})^\dagger \nabla_{\bm{k}} v(\bm{k})] \nonumber \\
    &= -i\nabla_{\bm{k}} \log \textrm{Pf}\,[v(\bm{k})].
\end{align}

We can easily prove that $\nabla \times [\mathcal{A}(\bm{k}) + \mathcal{A}(\bm{k}+\bm{k}_0)] = 0$,
so we choose a gauge so that $\mathcal{A}(\bm{k}) + \mathcal{A}(\bm{k}+\bm{k}_0) = 0$.
Thus, after the gauge transformation of a form $\textrm{Pf}\,[v(\bm{k})] \to \textrm{Pf}\,[v(\bm{k})]e^{-i\theta_{\bm{k}}}$,
we can fix a gauge to make $\textrm{Pf}\,[v(\bm{k})]=1$.
At the same time, the nontriviality coming from the definition of $\sqrt{\det}$ disappears.
The rest is to relate $v_{mn}(\bm{k})$ to $w_{mn}(\bm{k}) = \braket{u_{m,-\bm{k}}|\Theta_-|u_{n,\bm{k}}}$.
\begin{equation}
    w_{mn}(\Gamma_i) = -\braket{\psi_{m,\Gamma_i}|(\Theta_+ P)(-\tau^z)P|\psi_{n,\Gamma_i}}.
\end{equation}
This is because $P^2=-1$.  Using antilinearity of $\Theta$,
\begin{equation}
    w_{mn}(\Gamma_i) = -i\tau_n(\Gamma_i) \xi_n(\Gamma_i)v_{mn}(\Gamma_i),
\end{equation}
where $n$ includes an index $\tau$ and $\tau_n(\Gamma_i)$ is its $\tau^z$ eigenvalue.
First we note that
\begin{equation}
    \textrm{Pf}[w]^2 = \det[w] = \det[v] \prod_{n=1}^{2N} [-i\tau_n(\Gamma_i) \xi_n(\Gamma_i)].
\end{equation}
As we already saw $\xi_\alpha(\Gamma_i) = -\xi_\alpha(\Gamma_i+\bm{k}_0)$,
each pair of states with an opposite $\tau$ has an opposite sign.  Then,
\begin{equation}
    \textrm{Pf}[w] = \textrm{Pf}[v] \prod_{\alpha=1}^{N} [-i\xi_{\alpha}(\Gamma_i)].
\end{equation}
Thus, using $\textrm{Pf}[v] = 1$, a Fu-Kane invariant inside HBZ can be computed
from the product for two $k$-points $\Gamma_1$ and $\Gamma_2$.
\begin{equation}
    \delta = \prod_{i=1}^2 \textrm{Pf}[w(\Gamma_i)] = \prod_{\alpha=1}^{N} \prod_{i=1}^2 [-i\xi_{\alpha}(\Gamma_i)],
\end{equation}
which coincides with the previous definition for $N=2$.

In the final form, it is not apparent that this invariant is still valid when the
inversion symmetry is broken.  However, if we go back to the definition, the $\textrm{Pf}[w]$
is multiplied for every IIM inside HBZ, and thus the $Z_2$ invariant here can be defined solely
by the time-reversal symmetry.  We note that $\textrm{Pf}[w]$ has a meaning only after
identifying $(k_x,\,k_y)$ and $(k_x + \pi,\,k_y + \pi)$.

\bibliography{suppl}